# SAPHIR: A Pluricultural Authoring Tool to Produce Resources in Support of Education for Sustainable Development

Stéphanie Jean-Daubias[a]
*Univ. Lyon, UCBL, CNRS, INSA Lyon, LIRIS, UMR5205, F-69622 Villeurbanne, France*
*stephanie.jean-daubias@univ-lyon1.fr*

Keywords: Authoring Tool, Mobile and Informal Learning, Design Process, Sustainable Development.

Abstract: In this paper, we present SAPHIR, a multilingual authoring tool producing a Progressive Web App, usable on computers, tablets, and smartphones, online or offline. We presented our design process, the architecture of the system, the model on which it is based, and its main parts: SAPHIR it-self is the main software proposing activities to children to learn and play; MINE is the authoring tool used by pedagogical designers and resources translators to create and translate resources without requiring any programming skills; TAILLE is dedicated to teachers to whom he provides educational explanations to use SAPHIR with their learners. The different parts were used with both pedagogical designers and students.

## 1 INTRODUCTION

The sixth IPCC report (Intergovernmental Panel on Climate Change) draws alarming conclusions on the impacts of global warming, including dependence on fossil fuels and renewables, limitation of vital resources such as water. In particular, the IPCC urges developed regions to support developing countries, both socially and economically in facing environmental challenges. This collective awareness includes education. Indeed, building a more informed society will help tackling these challenges. However, there is still a long way to go to raise awareness among young people on these questions (IPCC, 2022).

Albatross Foundation (Albatross, 2023) addresses this issue. This non-profit organization was created in 2011 in Singapore, in 2012 in China, and in 2017 in France where it is recognized of public utility. Its aim is to educate for free young people to sustainable development in a scientific and fun way, thanks to innovative tools. Albatross has developed training programs and book donations on the environment: more than 15 000 children have been trained in China, France and Brazil using the "Train-the-Trainer" method, whether in small weakly connected villages or large hyper connected cities. An experiment on water purification is proposed for children aged 8 to 12 using the foundation's educational kit. Activities of the foundation are organized according to the classical elements (earth, water, air, fire) which were proposed to explain the nature (Laszlo, 2009).

The project presented in this paper takes place within the framework of a collaboration between the LIRIS computer science research lab and the Albatross Foundation. Its aim is to design, develop and test a multilingual and multicontext app for education to sustainable development.

In this paper, we present SAPHIR (education to Sustainable development with Albatross through Pedagogical activities: Humans Involved in the Respect of their environment), an environment consisting of different parts (Saphir, 2023): SAPHIR itself is the main software proposing activities to children to learn and play; MINE is the authoring tool used by pedagogical designers and translators to create and translate resources without requiring any programming skills; TAILLE is dedicated to teachers to whom he provides educational explanations to use SAPHIR with their learners. In the following, we will introduce the issues of our work and the associated state of the art. After a presentation of the roles of the various actors involved, we will show the architecture of our system and the model that we propose. We will then come back to the design process that we have adopted, and expose some of the tests conducted with the app. The article will end with a conclusion and a discussion.

[a] https://orcid.org/0000-0003-2209-6376

# 2 ISSUES, STATE OF THE ART AND CONCEPTION CHOICES

Depending on the context, the association needs a light application working even offline on computers, tablets, and smartphones, for classrooms and informal learning, available in different languages. But it also requires offering the possibility to add news contents, to translate these contents quickly and flexibly, according to the deployments planned by the association.

## 2.1 Issue and Research Questions

The general issue of this project is thus to conceive an evolutive and flexible application suited to different contexts and cultures. We identified three main research questions while treating this issue: how to allow non-computer scientists to create contents for this app (Q1); how to consider the multilingual and multicultural context of the project (Q2); and, how to take into account the different contexts of use of the app (Q3).

## 2.2 State of the Art and Choices

For our first research question (Q1: how to allow non-computer scientists to create contents), the use of authoring tools seems to be an interesting direction.

Therefore, research in TEL (Technology Enhanced Learning) identifies classically three ways to create pedagogical resources (Jean-Daubias et al., 2009): fully automatic resources generators, semi-automatic generators, and manual generators (also called authoring tools).

Fully automatic generators, see for example (Burton, 1982), are complex programs specific to a dedicated domain. They allow to produce resources in this domain without any intervention of the pedagogical designer. This approach is very efficient as it can produce rapidly many resources without any effort, but do not allow a personalisation of the produced resources. Another disadvantage of this approach is that it is complicated and long to conceive and implement. It does not suit our needs, as we want to define our resources precisely to ensure that they will be adapted to our different contexts.

Semi-automatic generators, see for example (Jean-Daubias and Guin, 2009), are close to automatic generators, except that they let more place to pedagogical designers to suit their needs: users can intervene in the resources creation process by specifying a set of constraints on them in order to obtain results more suited to their needs, still in the scope of the generator.

Manual generators (commonly called authoring tools), see for example (David et al. 1996) and (Van Joolingen and De Jong, 2003), guide the pedagogical designers step by step in the resources design and gives them a great liberty. Limits of this approach is that it is time consuming for the user, as he must precisely define each exercise and their solutions. In the context of our project, the advantages of this approach make its disadvantages acceptable.

*We have therefore chosen an approach based on an **authoring tool** to enable non-computer scientists' educational designers to create educational resources on sustainable development.*

For our second research question (Q2: how to take into account the multicultural context of the project), a resource designed in one language cannot always be translated directly into another: it may require adaptation to the target culture (for example, the operating rules for recycling waste vary according to countries or regions). We decided to establish different versions of the resources: each version is not only a translation of the original resource but can also include adaptations to the cultural context of the concerned country or region.

*So, we created a **multilingual authoring tool**, adding to the classical user roles of such tools (administrator and educational designers) a **translator role**, for people in charge of adapting to different cultures the various resources produced by educational designers.*

For our third research question (Q3: how to take into account the different technical contexts of use of the app), we need to propose a flexible application easy to update and to use in different technical contexts. In particular, the application must be usable offline, in places where learners won't have an easy access to the internet. The app also needs to be able to connect to the internet to allow updates in particular to add new contents. The app also needs to be available on various devices (computers, tablets, and smartphones), depending on the devices available in the different use contexts. A good solution to deal with this triple need is a PWA (progressive web app): a type of application software delivered through the web, working on various operating systems with a standard-compliant browser, including desktop and mobile devices (Mozilla, 2023). We like to present this type of website to the general public as "a website disguised as a mobile app".

*Thus, we created a **multilingual (Q2) authoring tool (Q1) producing a Progressive Web App (Q3),***

*usable on computers, tablets, and smartphones, online or offline.*

We considered that using an LMS (Learning Management System) is too rigid to fully meet our needs, especially to create an app clearly identifiable linked with Albatross Foundation.

# 3 ARCHITECTURE AND RESOURCES MODEL

Figure 1 shows the general architecture of SAPHIR. The system consists of four main parts: SAPHIR itself is the main software proposing to children activities to learn and play; MINE is the authoring tool used by pedagogical designers and resources translators to create and translate resources; TAILLE is dedicated to teachers to whom it provides educational explanations; finally, the database stores the resources associated with the different modules. In the following, after a presentation of the roles of the various actors involved, we will present this architecture and its different parts.

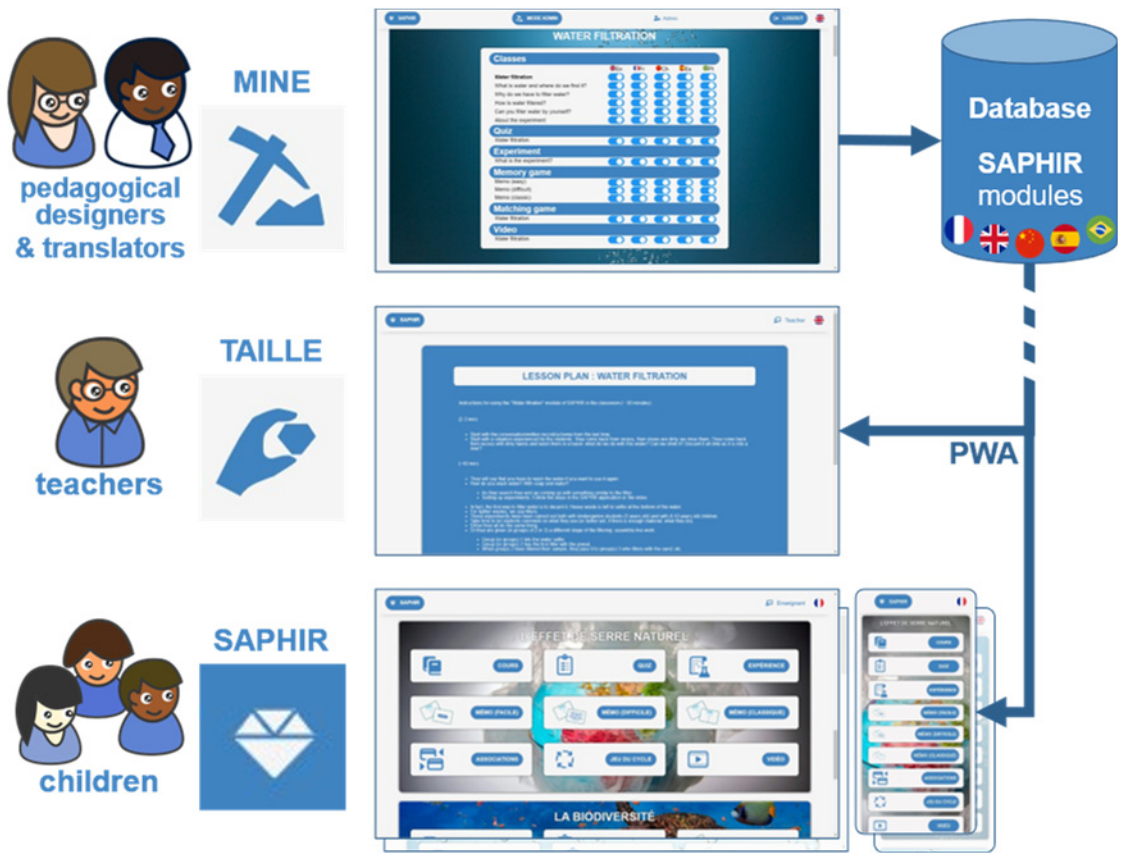

Figure 1: Architecture of SAPHIR environment.

## 3.1 Actors

SAPHIR environment involves users of various roles.

The **developers** are computer scientists, they created and still enrich the environment (MINE, TAILLE and SAPHIR itself).

The **administrator** manages the environment.

The **educational designers** are experts in sustainable development, teachers or didacticians, they use MINE to create new resources.

The **translators** are experts in sustainable development in the target language, they also need to have a good knowledge of the context in which SAPHIR will be used in the targeted country. They use MINE to translate resources to make them available in new contexts.

The **teachers** are usually prescribers of SAPHIR to their pupils and also users of TAILLE in the aim of studying the way to use SAPHIR's resources in their classroom and more generally with their pupils.

The **children** are SAPHIR final users, they use it to learn concepts linked to sustainable development.

## 3.2 Architecture

SAPHIR is developed as a Progressive Web App (PWA) associated with a mySQL database. It allows SAPHIR and TAILLE to work even offline on computer, tablets, and smartphones after the application had been installed locally. However, MINE, the authoring tool, cannot be available offline as it requires an internet access to supply the database. The access to SAPHIR is open, the access to TAILLE too, but it requires to activate the teacher mode, while the access to the authoring tool requires a connection with a login and a password.

The names of SAPHIR, TAILLE and MINE are based on a gem metaphor in French: *saphir* is the French translation for sapphire; a *mine* is the place where we find gems; *taille* is the French word for *cut* of gems, which is the operation aiming at transforming rough gems into more usable ones. So, the metaphor is that the educational designers find rough gems in the MINE, teachers adapt them to their use in TAILLE, and pupils finally use the refined gems in SAPHIR.

### 3.2.1 MINE: An Authoring Tool for SAPHIR

MINE (Means to Integrate New Elements in SAPHIR) is SAPHIR authoring tool, usable without any programming skills. This is the interface where pedagogical designers can create interactively new modules and news activities (lessons, quiz, small games). It is also the place where translators can add new translations of existing modules and their resources. MINE also allows the administrator to add new languages. Thus, MINE allows non-computer scientists to create new contents for SAPHIR.

The MINE of SAPHIR includes a content creation interface allowing the creation of new contents (figure 3 shows the creation of a lesson page with a labelled picture, associated with one question of the quiz previously created in MINE) and their translations (figure 2 shows as an example the translation of a lesson page from English to Chinese and Spanish in MINE).

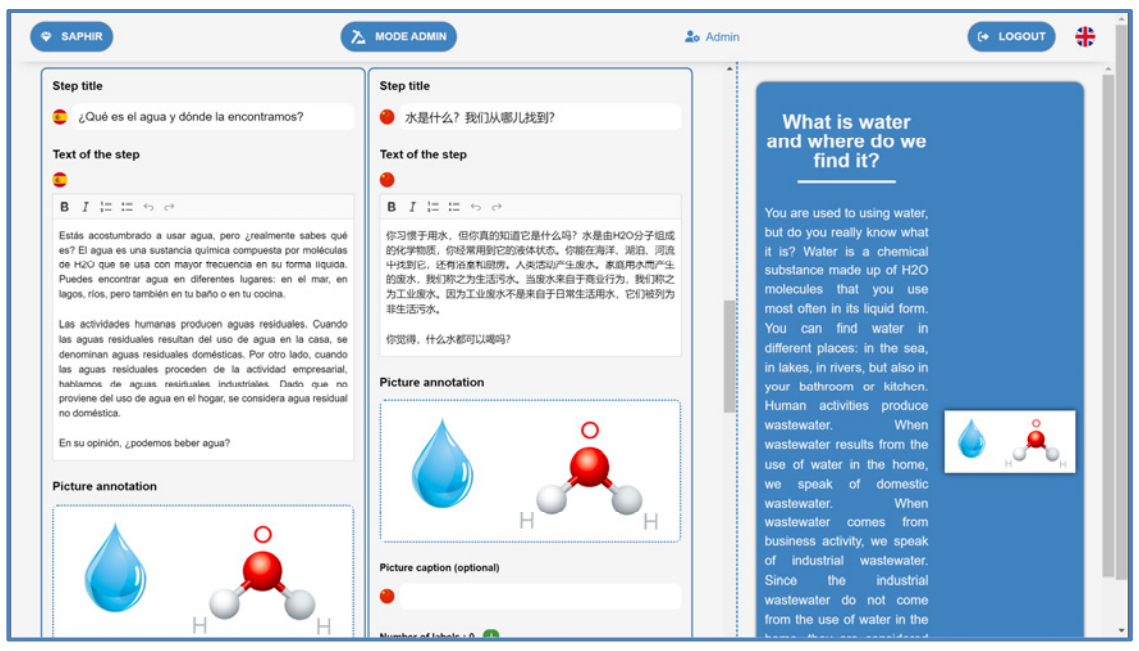

Figure 2: Screenshot of MINE during the translation of a lesson from English (right) to Spanish and Chinese (left).

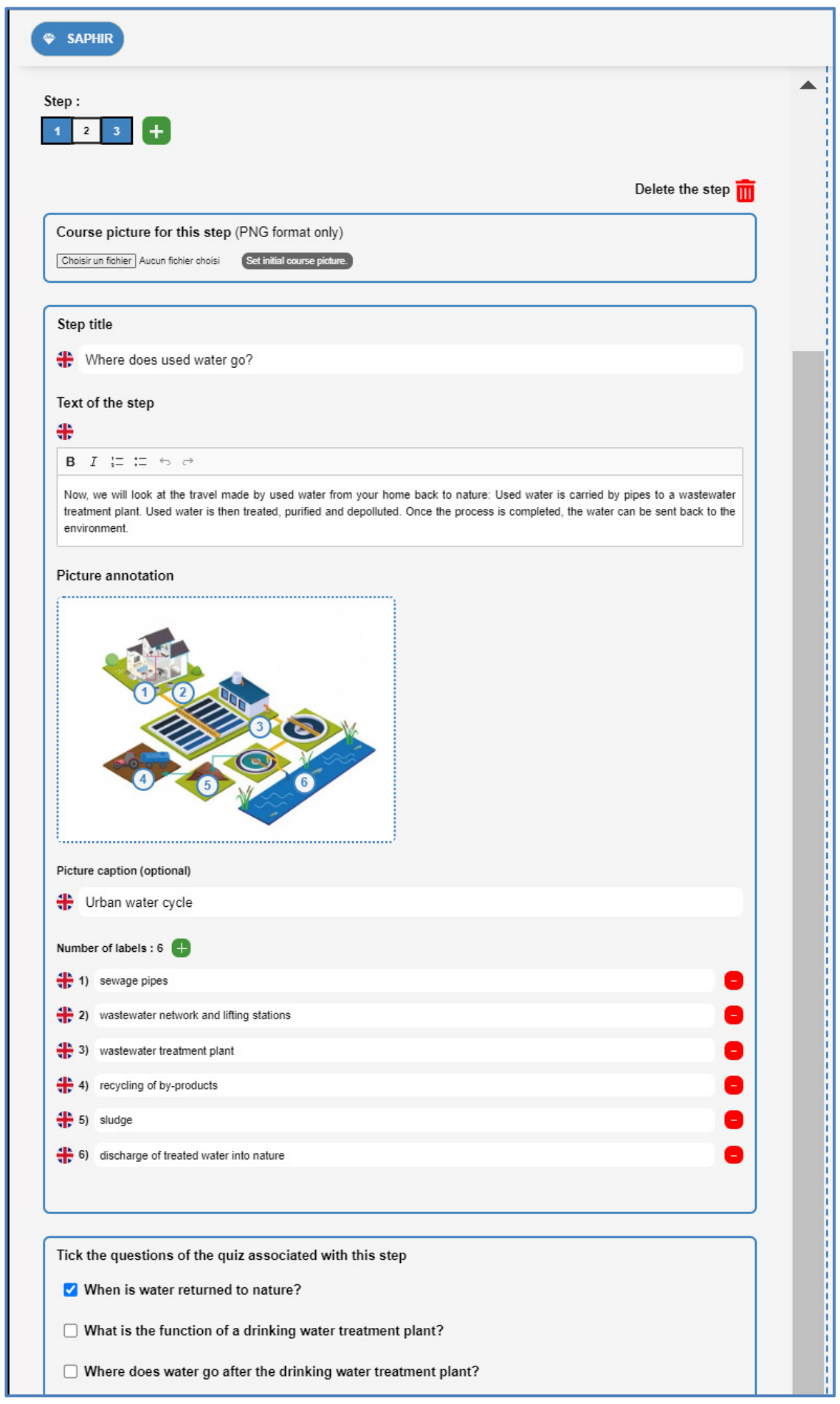

Figure 3: Screenshot of the creation of a lesson with MINE.

### 3.2.2 SAPHIR Resources Model

Creating a new module with MINE consists of completing the different parts of the module model. This model is shown in figure 4 and described below.

A **Module** consists of a lesson, a quiz, 1 to 3 memo-games, an association-game, a cycle-game, an experiment, a short video, and the associated pedagogical-support (available only if logged-in as teacher in TAILLE). This general model ensures homogeneity between modules of SAPHIR.

**Module** = (Course; Quiz; Memo-Game; Association-Game; Cycle-Game; Experiment; Video; Pedagogical-Support)

**Quiz** = *n* Question, $n \geq 1$

**Question** = (Title; *m* Q-Propositions; Explanation), $m \in [2, 10]$

**Q-Propositions** = (Title; Personalized-Explanation; Validity)

**Course** = *p* Page

**Page** = (Title; Text; Picture; Caption; *r* Tag; *q* Quiz), $q \leq n$

**Tag** = (Text; Coordinate-Horiz; Coordinate-Vert)

**Memo-Game** = 6 (Picture; Title; Definition)

**Association-Game** = (2 Category; *t* A-Proposition)

**A-Category** = (Title, Picture)

**A-Proposition** = (Title; Category; Personalized-Explanation)

Figure 4: SAPHIR educational resources model.

A **Quiz** is defined as a set of questions. Each question includes a title, various propositions (between 2 and 10) and a general explanation in case of error of the learner). The propositions (correct or not, which are modelled by the validity) are represented by a title and an optional personalized explanation, which allows to give better feedback to the learner.

A **Lesson** consists of different pages. Each lesson page comprises a title, the text of the lesson page, possibly a picture, its caption, and optional tags to explain the illustration. A page is linked with the suited questions of the module quiz. Indeed, in order to propose more interactive lessons, each lesson page propose a quiz to allow the learner to test himself after reading the content (cf. figure 3).

The lesson can be illustrated to be more suited to young children and more generally to propose easy-to-access and enjoyable resources. Each picture can be captioned and added with numbered tags: the numbers are located on the picture (location defined by the horizontal and vertical coordinates of the tag) and the associated text is displayed to the learner when he clicks on the tag on the picture. This functioning (cf. figure 5) has a triple advantage: it makes the user more active in his learning in a constructivist approach (Piaget, 1971); it facilitates the translation of the tags (which are text, out of the picture); and it enhance the numerical accessibility of SAPHIR and his compatibility with accessibility tools, for example the narrator used by blind people (WCAG, 2008).

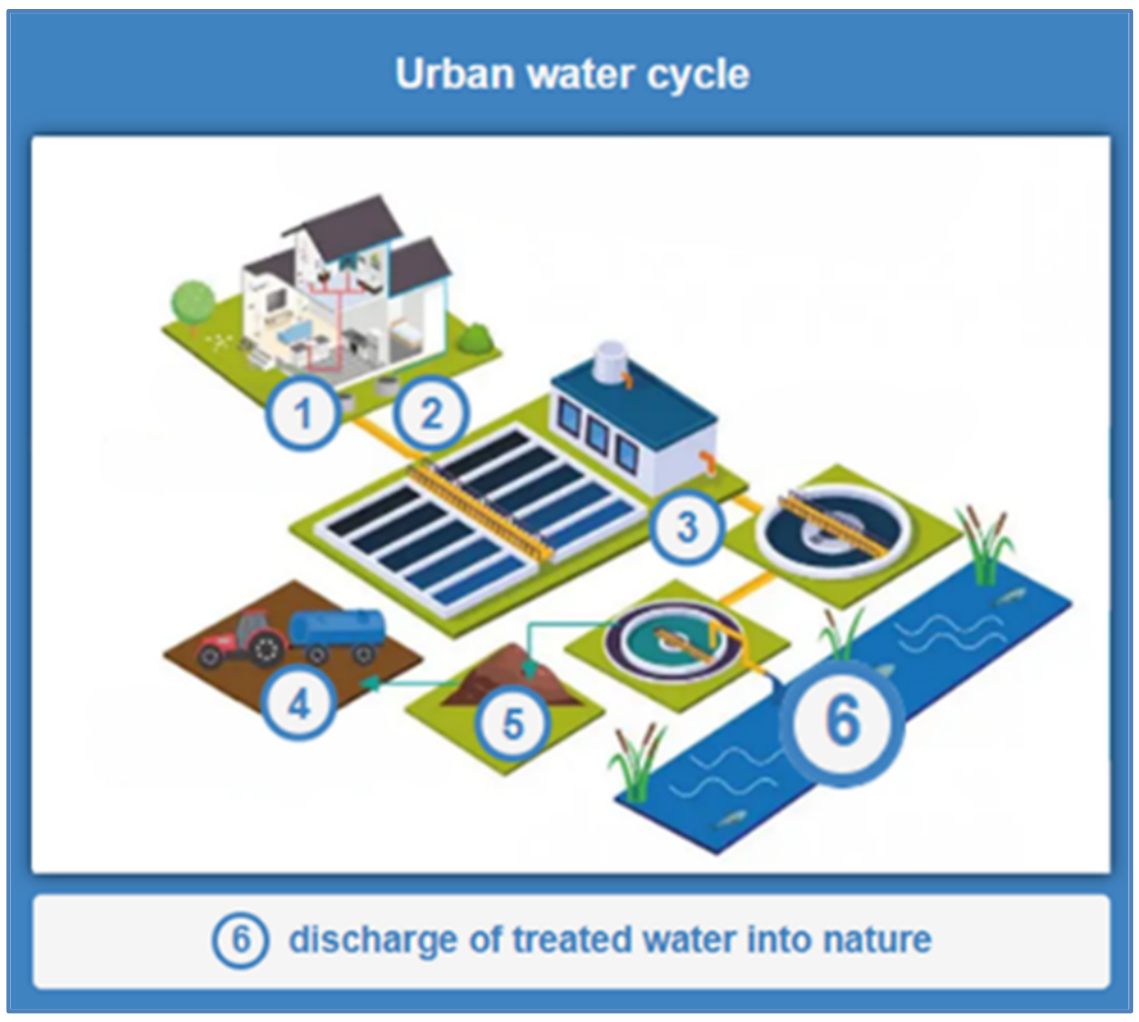

Figure 5: Interactive picture in SAPHIR.

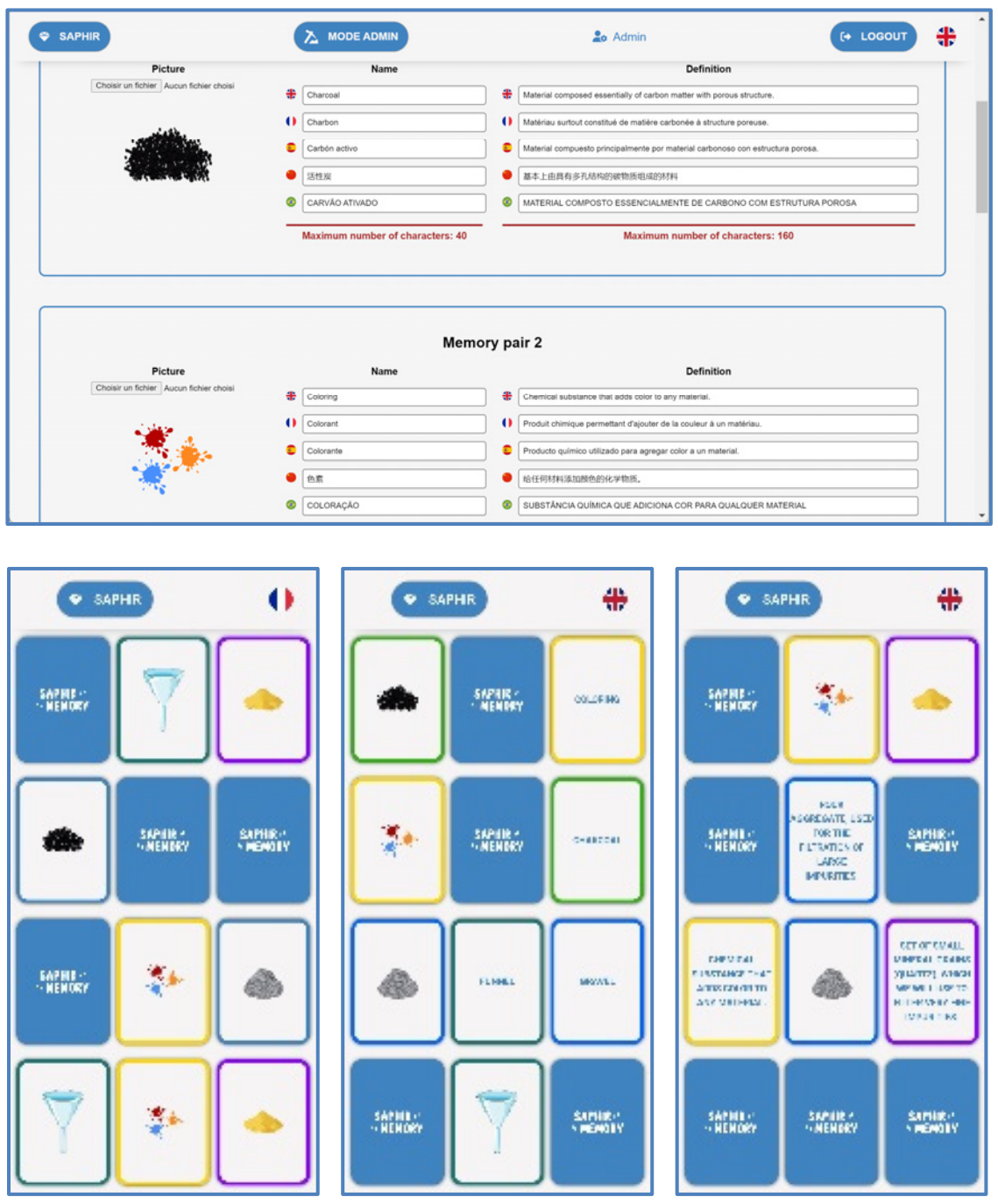

Figure 6: Screenshots of MINE during the creation of a memory game and of 3 types of memory games in SAPHIR.

A **Memo Game** model is a set of 6 triplets associating a picture to a title and a definition. This model allows the creation of three different memory games: the "classical memory" requires the child to found two identical pictures; the "easy memory" requires associating an image to the corresponding word, in order to work on the lesson's vocabulary; and the "difficult memory" asks to associate the name of the concept to its definition (cf. figure 6). The number of triplets is limited to 6 to ensure a comfortable use and a correct display even on mobile phone, in respect with usability considerations (Nielsen, 1994) (Tricot, 2007).

The **Association Game** (cf. figure 7) asks learners to associate different concepts to two opposite categories. An association game is modelled as a set of 2 categories associated with various propositions. The learner will have to associate each proposition to the right category, linked with the lesson. A category is defined by a title and a picture representing it. Each proposition contains a title, the matching category and a personalized explanation used to explain his mistake to the learner in case of error.

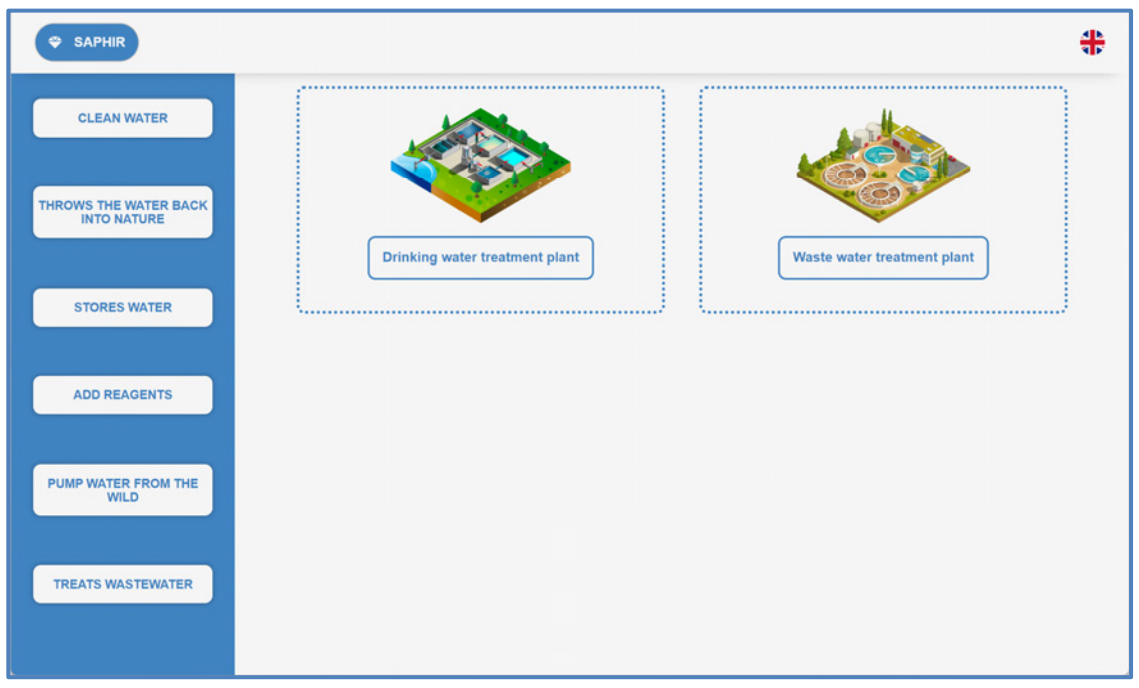

Figure 7: Association game in SAPHIR.

MINE also allows the pedagogical designers to add a link to a pedagogical **video**.

Other resources of SAPHIR (**experiment**, cf. figure 8 and **cycle game**, cf. figure 9) are not available in MINE for now. The addition of such resources requires developments by computer scientists.

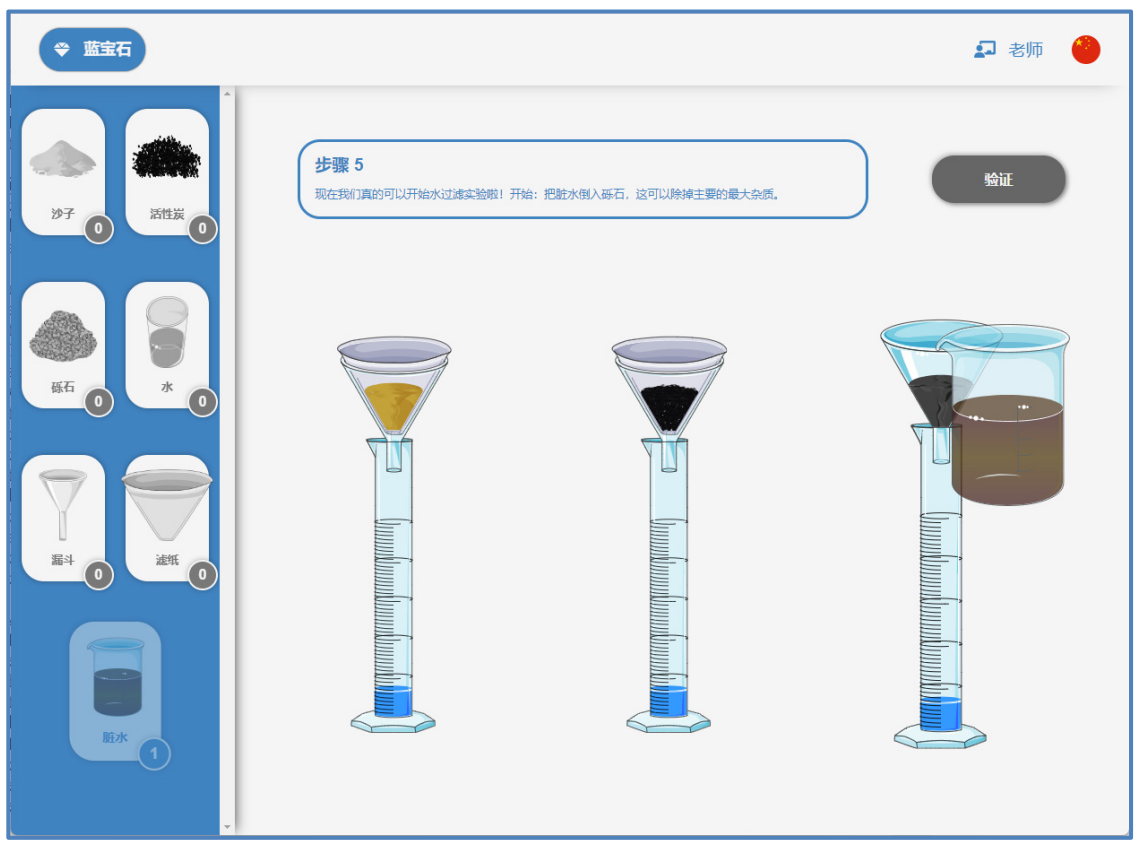

Figure 8: Water filtration experiment in Chinese.

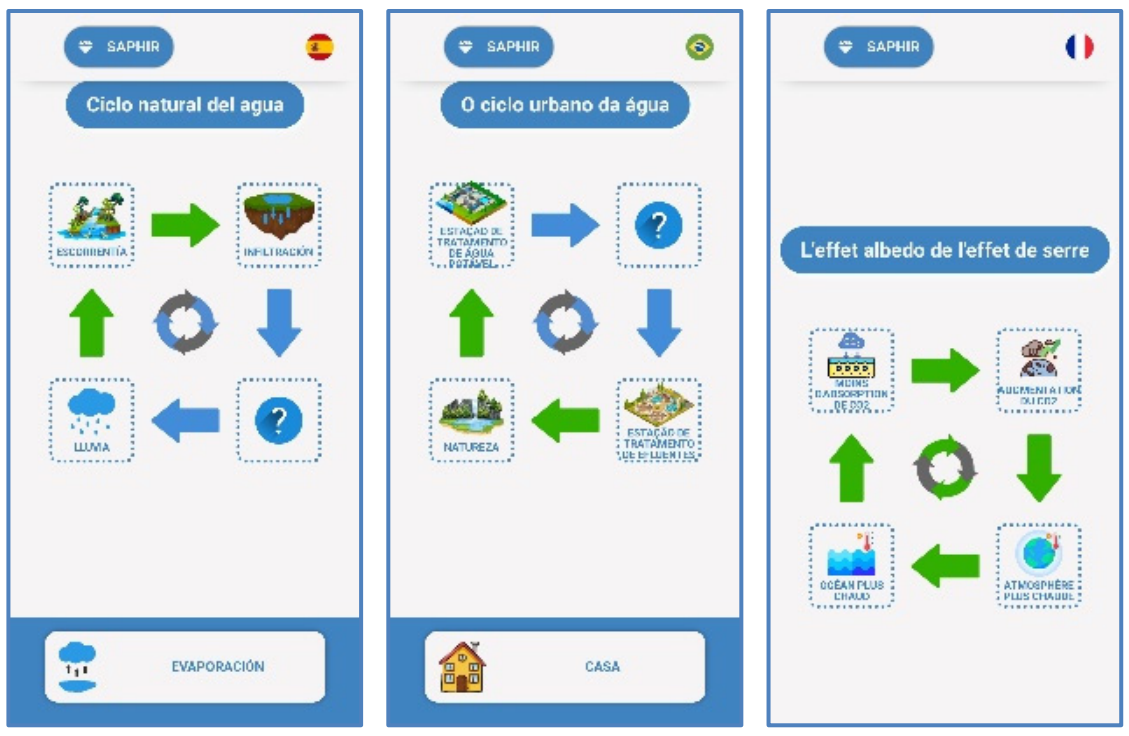

Figure 9: Cycle games in SAPHIR (natural water cycle in Spanish, urban water cycle in Portuguese, and greenhouse effect in French).

### 3.2.3 TAILLE: The Teacher Interface for SAPHIR

TAILLE (TrAiners Interface to Learn in different Languages how to teach Environmental issues with SAPHIR) is SAPHIR teachers' interface. It consists of SAPHIR enhanced with a pedagogical description of each module (cf. figure 10). The aim is to provide teachers with explanations on how they can use these resources in their teaching or more generally with their pupils.

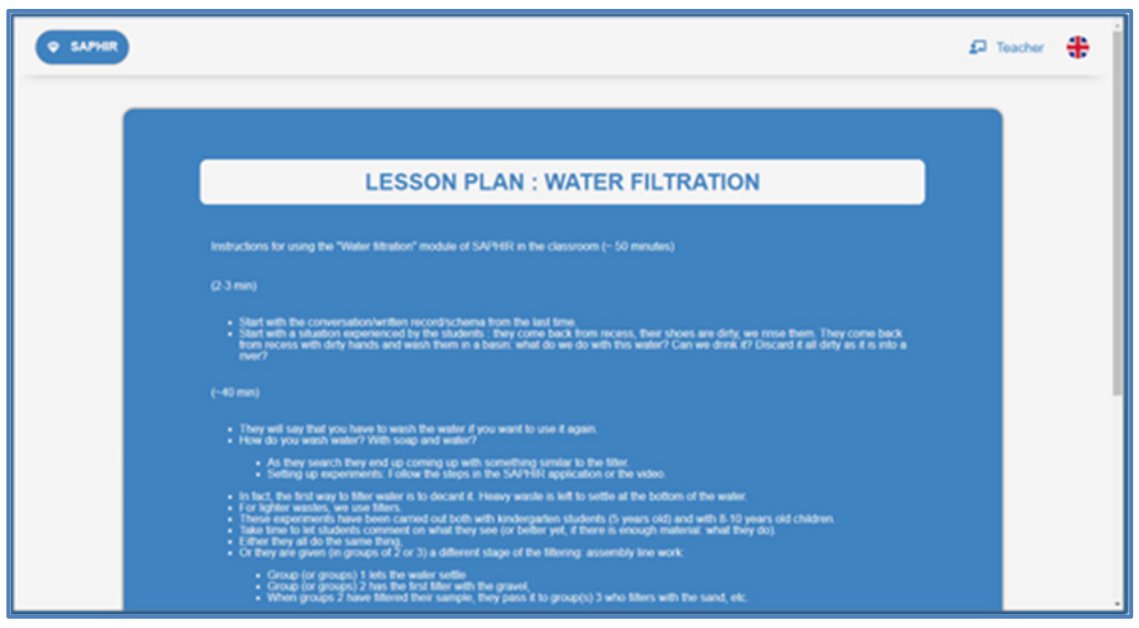

Figure 10: Screenshot of TAILLE, the teacher interface.

### 3.2.4 SAPHIR: A Multilingual App for Education to Sustainable Development

SAPHIR itself is the learners' interface of SAPHIR environment. It currently includes six modules (water filtration, urban water cycle, natural water cycle, the greenhouse effect, the natural greenhouse effect, biodiversity), each module is assigned to one of the four categories: water, air, earth, and energy, based on the traditional Chinese representation of the natural elements used by Albatross Foundation (Laszlo, 2009). Each module contains up to nine resources (among lesson, quiz, 3 types of memory games, association game, cycle game, experiment, video, as defined in the model presented in section 3.2.2). The resources are currently available in up to five languages (English, French, Chinese, Spanish, and Brazilian-Portuguese) chosen according to the requirements of the Albatross Foundation activities. 43 different resources are available for the moment.

We made the choice not to rely on internet connexions, in order to propose a simple system, without technical constraints for the learners that do not require personal data management. As it is a tool for informal learning, the resulting lack of learner profile is not considered as a problem.

We decided to give access to the different modules and resources freely, that is to say without any constraint of order, to allow the learner, or the teacher, to choose the activities he wants to follow. The idea is to facilitate access to knowledge on sustainable development, in an informal and flexible way. Thus, there is no pedagogical scenario to link the different activities, except between lessons and quizzes: each lesson page is proposed with a suited question taken from the corresponding quiz module to help the learner to check if he has understood the lesson. When the learner launches a quiz, as a set of five questions is generated semi-randomly from the available questions, covering the scope of the different lesson pages and taking into account the questions already answered during the lesson.

The resources are linked with the activities of Albatross Foundation. For example, the water filtration module is linked to the real water filtration experiment proposed by the foundation to sensitize hundreds of children in the importance of water and the fragility of its quality; the biodiversity module is linked to the Servan's garden initiative of the foundation, a permaculture pedagogical garden to develop sustainable agriculture, respecting the environment, promoting biodiversity to sensitize and educate young people to "grow and eat well" (Albatross, 2023).

## 4 DESIGN AND EVALUATION PROCESS

### 4.1 Design Methodology

For the design and the development of SAPHIR environment, we adopt a user-centred method (Norman and Draper, 1986) with an iterative approach. We designed personae for each role of users (admin, educational designer, translator, teacher, learner), in order to identify easily the different target publics.

Figure 11 shows the three main stages of SAPHIR design process.

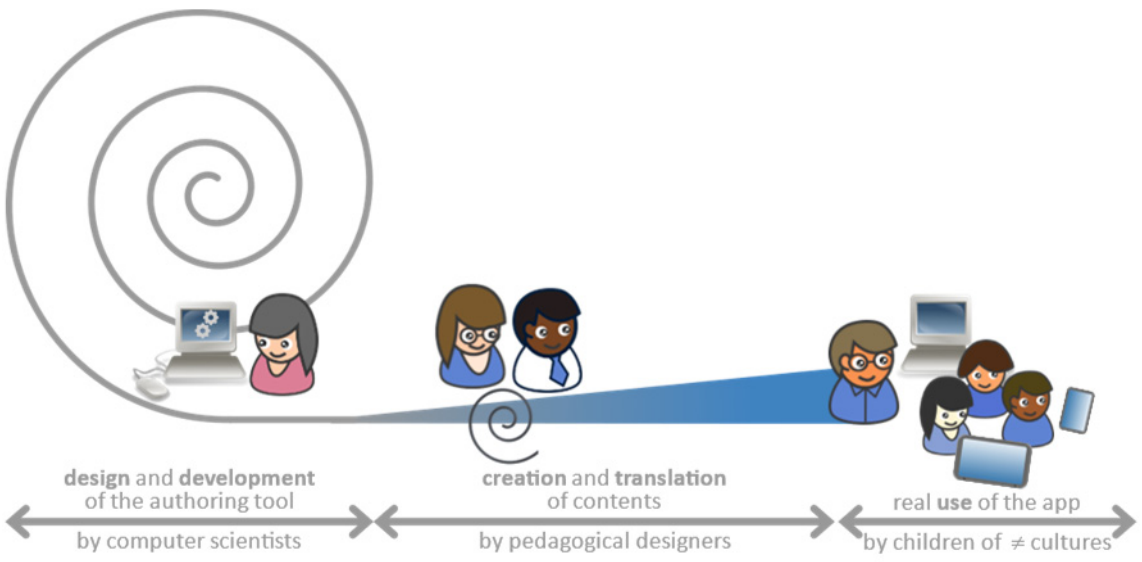

Figure 11: Diagram of SAPHIR design process.

The first stage of the process (left part of figure 11) corresponds to the iterative **design and development of the global environment** by *computer scientists*. Our iterative approach (symbolized by the left spiral in the figure) enabled us to propose a preliminary version of the app to communicate with the association members and to conduct early tests with children, but also teachers. Then, the authoring tool was progressively developed, enabling to create more easily new contents. At this stage, SAPHIR contained only three modules on water and was available only in English, French and Chinese; as for MINE, it was an empty shell.

Once the environment and the authoring tool had been sufficiently developed, the second stage (central part of figure 11), consists of **filling** by *experts* and *teachers*, including *translators* with scientific knowledge, **this empty shell with the educational contents** expected by the association. This enrichment of SAPHIR by the addition of pedagogical contents is symbolized in the figure by the blue bar which widens progressively. SAPHIR has thus grown from 3 to 6 modules, now gathering 43 resources available in several languages.

The third stage (right part of figure 11) is the **use** of the various resources of SAPHIR by *children* from different cultures, in different languages, with or without the support of their *teachers*.

*We consider that contents creation is part of the design of the system: pedagogical designers complete the design by their uses of the authoring tool.*

## 4.2 Content Creation and Translation

The creation of the contents, associated with one of the classical elements (earth, water, air, fire), is a collaborative and iterative work involving at different stages scientists, experts, teachers, and translators. Each person involved in the process can modify the contents according to his skills.

## 4.3 Use of SAPHIR in Different Contexts

MINE has been successfully tested with pedagogical designers: the interface allows non-computer scientists to create and translate new resources.

SAPHIR including the resources produced by pedagogical designers with MINE was tested several times with learners in different contexts. In the following, we present 3 of these tests chosen for their diversity. The first test was conducted in France in June 2020, in French language, it involved 6 pupils between 6 and 11 years old, in a Covid-scholar context in a small country school (cf. bottom right picture of figure 12). The app was used collectively on the classroom interactive whiteboard. The teacher and the pupils appreciated the app. The test validated the usefulness of the app and allowed to identify some usability problems (Tricot, 2007) which were subsequently corrected. Alongside other assessments in France, another test was conducted in May 2021 in China, with young pupils, on computers, tablets, and smartphones (cf. 3 top pictures of figure 12) in Chinese. This test validates the usability on different devices and in Chinese language. The most recent test took place near San Paolo in Brazil in November 2022 with 30 pupils, in Portuguese (cf. bottom left pictures of figure 12: students using the water filtration kit). The test showed the interest of using SAPHIR with older pupils and validated the Portuguese version of the app.

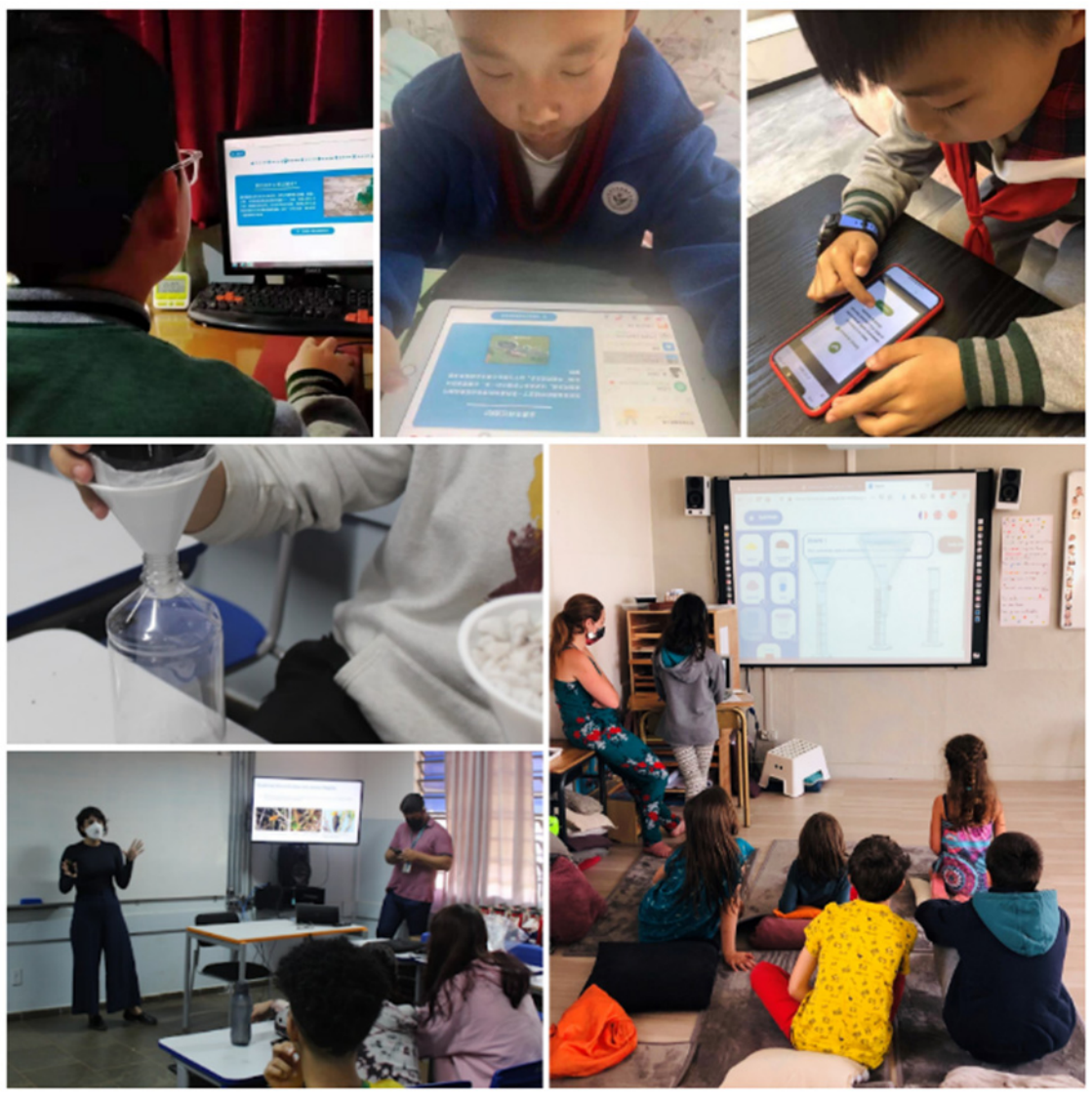

Figure 12: Use of SAPHIR in classrooms.

Apart from these tests, SAPHIR is also used independently of classes. We identified more than 20 000 hits from computers all over the world.

# 5 CONCLUSIONS

## 5.1 Discussion

The fact that non-computer scientists can create and translate new contents with the MINE of SAPHIR is a first validation step for our authoring tool. But we need to continue evaluations to carefully validate MINE.

As for the contents created, after being validated by teachers and experts, they were used with several children of various languages and cultures, in different contexts. Both teachers and students found their use of SAPHIR useful and enjoyable.

## 5.2 Conclusion

In this project, we addressed the issue of how to conceive an evolutive and flexible application suited to different contexts and cultures. It conducted us to answer three questions: how to allow non-computer scientists to create contents for this app; how to consider the multicultural context of the project and the different contexts of use of the app.

As an answer to these questions, we created a multilingual authoring tool producing a Progressive Web App, usable on computers, tablets, and smartphones, online or offline.

In this paper, we presented our design process, the architecture of the produced system, the model on which it is based, and the main parts of the system, including MINE, the authoring tool, TAILLE, the teachers' interface, and SAPHIR itself (cf. figure 13), the children app. The different parts were tested successfully.

Several perspectives are open to us now. First of all, we will continue adding new contents and new translations to enlarge the scope of the app to cover more broadly the themes of sustainable development addressed by the Albatross Foundation in order to sensitize more children to sustainable development. Moreover, new languages could be added in case of deployment of the app in new countries. In addition, we are planning uses of the app in a new context: scientific mediation events with the general public.

We are also considering working with children on the next resources with the idea of proposing "an app for children made by children". For this, we have undertaken projects with several classrooms in different contexts. The idea is to make pupils working together in classroom to establish new modules to add to SAPHIR for future young users. Additionally, we are going to propose to future teachers and didacticians to use MINE to make them work on content creation in a didacticians approach.

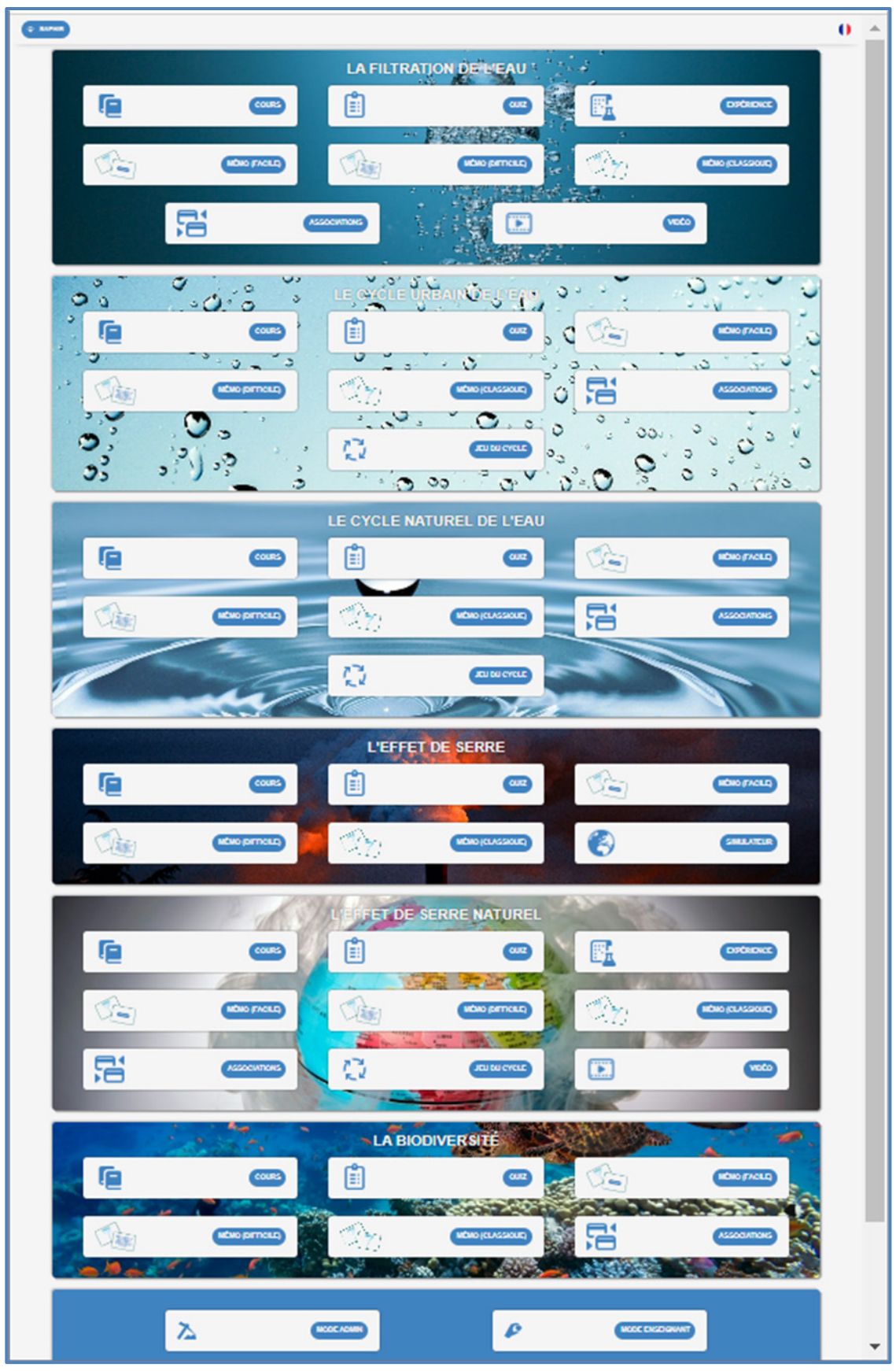

Figure 13: Screenshot of SAPHIR's main screen.

# ACKNOWLEDGEMENTS

SAPHIR app was awarded the Franco-Chinese Teams Innovation Award of MEDEF organisation in 2019, and the Belgian King Baudouin Foundation Award in 2020 and 2022.

The app was developed by students from the computer science master's degree of Lyon 1 University: especially Lucas, Mikael, Vichith, Jonathan, Dorian, and Dorian, thanks to them. The author also thanks Philippe Daubias from ENS de Lyon for administrating the SAPHIR's web server and all the people who participated in the creation or translation of the contents, especially Albatross team and her founder Ghislaine Bouillet-Cordonnier.